\documentclass[sigconf]{acmart}
\AtBeginDocument{  \providecommand\BibTeX{{    \normalfont B\kern-0.5em{\scshape i\kern-0.25em b}\kern-0.8em\TeX}}}

\copyrightyear{2022} 
\acmYear{2022} 
\setcopyright{acmcopyright}\acmConference[EuroMPI/USA'22]{EuroMPI/USA'22: 29th European MPI Users' Group Meeting}{September 26--28, 2022}{Chattanooga, TN, USA}
\acmBooktitle{EuroMPI/USA'22: 29th European MPI Users' Group Meeting (EuroMPI/USA'22), September 26--28, 2022, Chattanooga, TN, USA}
\acmPrice{15.00}
\acmDOI{10.1145/3555819.3555825}
\acmISBN{978-1-4503-9799-5/22/09}

\usepackage{algorithmic}
\usepackage[algo2e,ruled,vlined]{algorithm2e}
\SetCommentSty{scriptsize}
\SetKwComment{tcc}{\{}{\}}
\SetKwIF{If}{ElseIf}{Else}{if}{}{else if}{else}{endif}
\SetKwFor{While}{while}{}{endw}
\SetKwFor{ForEach}{for each}{}{endfch}
\newtheorem{exmp}{Example}[section]

\begin{document}

\title{A Locality-Aware Bruck Allgather}

\author{Amanda Bienz}
\affiliation{  \institution{University of New Mexico}
  \city{Albuquerque}
  \state{New Mexico}
  \country{USA}}
\email{bienz@unm.edu}
\orcid{0002-8891-934X}

\author{Shreeman Gautam}
\affiliation{  \institution{The University of Utah}
  \city{Salt Lake City}
  \state{Utah}
  \country{USA}}
\email{u1092041@utah.edu}

\author{Amun Kharel}
\affiliation{  \institution{Virginia Tech}
  \city{Blacksburg}
  \state{Virginia}
  \country{USA}}
\email{akharel@vt.edu}

\renewcommand{\shortauthors}{Bienz et al.}

\begin{abstract}
Collective algorithms are an essential part of MPI, allowing application programmers to utilize underlying optimizations of common distributed operations.  The MPI\_Allgather gathers data, which is originally distributed across all processes, so that all data is available to each process.  For small data sizes, the Bruck algorithm is commonly implemented to minimize the maximum number of messages communicated by any process.  However, the cost of each step of communication is dependent upon the relative locations of source and destination processes, with non-local messages, such as inter-node, significantly more costly than local messages, such as intra-node.  This paper optimizes the Bruck algorithm with locality-awareness, minimizing the number and size of non-local messages to improve performance and scalability of the allgather operation.
\end{abstract}



\maketitle

\section{Introduction}
Parallel architectures are continually advancing, with current state-of-the-art systems achieving exascale performance.  However, parallel applications often fail to take full advantage of available compute power on these systems due to communication constraints.  Collective algorithms, provided as part of the Message Passing Interface (MPI), optimize common distributed algorithms, allowing application programmers to utilize existing optimizations.  

Commonly used collective algorithms include reductions, such as summing together a distributed array of data, broadcasting data from a single process, and gathering data onto a single process.  Furthermore, variations of these, such as the all-reduce and all-gather, broadcast the result across all active processes.  The underlying implementations for collective algorithms typically rely on minimizing the message count for small data sizes and message size for larger amounts of data, to optimize the overall cost.  The Bruck algorithm~\cite{Bruck}, for example, achieves an optimal $\log_{2}(p)$ message count, where $p$ is the number of processes.  However, there is no accounting for the cost of each individual message throughout this algorithm.  Non-local messages, such as those that are injected through the network, are typically more costly than local, or intra-node, communication.  Therefore, the Bruck algorithm can be further optimized by exchanging non-local communication for additional local messages.

In this paper, we present a novel locality-aware Bruck algorithm, which minimizes the total number of non-local messages communicated by any process, while also reducing the amount of non-local data.  The remainder of the paper is outlined as follows.  Section~\ref{sec:background} describes existing algorithms for all-gather operations with small data sizes and provides an analysis of the locality of each required step.  A locality-aware implementation of the Bruck MPI\_Allgather algorithm is presented in Section~\ref{sec:locality}.  Performance models for both existing and locality-aware all-gathers are provided in Section~\ref{sec:perf_models}, and performance results are presented in Section~\ref{sec:results}.  Finally, Section~\ref{sec:conclusions} provides concluding remarks.

\section{Background}~\label{sec:background}
The MPI\_Allgather is initialized on an array of $m$ values, which are evenly distributed so that each of the $p$ processes holds $\frac{m}{p}$ unique values of the array.  When the operation returns, each process holds all $m$ values of the originally distributed array.  The cost of an all-gather operation can be estimated with the postal model
\begin{equation}
    T \leftarrow \alpha \cdot n + \beta \cdot s~\label{eqn:postal}
\end{equation}
where $\alpha$ is the per-message latency, $\beta$ is the per-byte transport cost, $n$ is the number of messages communicated by any process and $s$ is the total number of bytes sent from a single process.  
\begin{figure*}[ht!]
    \centering
    \includegraphics[width=0.7\textwidth]{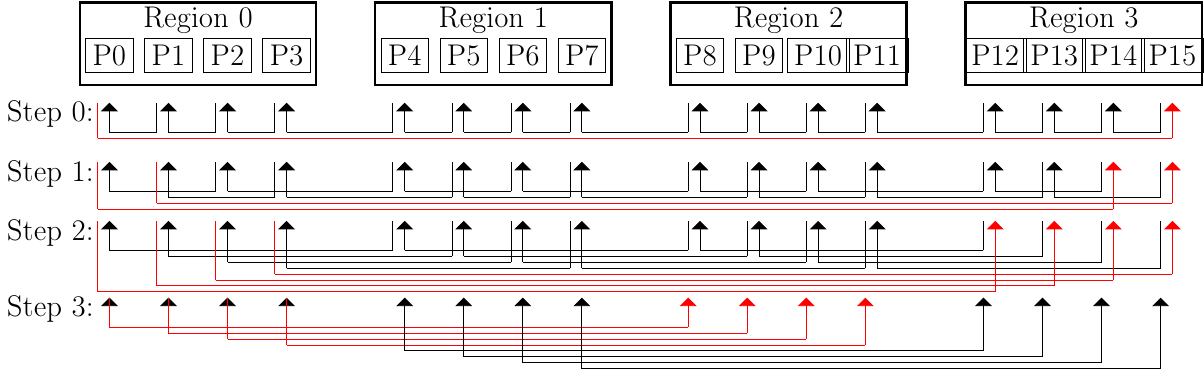}
    \caption{The communication pattern for each step of the Bruck allgather algorithm for Example~\ref{exmp:allgather}.  The red arrows highlight non-local communication originating from any process in region $0$ during each step of the algorithm.}
    \label{fig:bruck}
\end{figure*}

Throughout this paper, various all-gather routines will be examined for the process count and data as described in Example~\ref{exmp:allgather}.
\begin{exmp}\label{exmp:allgather}
    Assume there are $16$ processes, each containing a single value equivalent to its process id.  For example, process $P0$ is initialized with the value $0$, process $P1$ is initialized with the value $1$, and so on.  Furthermore, assume that groups of $4$ processes are grouped into a region of locality, so that communication within each region is less expensive than communication between regions.  Assume an all-gather is performed on this data, so that after the operation, all processes hold an array containing the values $0$ to $15$.
\end{exmp}

The all-gather operation is implemented in a variety of ways.  Recursive-doubling, dissemination~\cite{dissemination}, and the Bruck~\cite{Bruck} all-gather are all tree-like algorithms with $\log_{2}(p)$ steps.  
\begin{algorithm2e*}[ht!]
  \DontPrintSemicolon  \KwIn{$\texttt{Comm}$\tcc*{MPI Communicator}
        $\texttt{id}$\tcc*{Process ID in Communicator}
        $p$\tcc*{Number of Processes in Communicator}
        $\texttt{init\_data}$\tcc*{Initial data to be gathered} 
        $n$\tcc*{Number of values in $\texttt{init\_data}$}
        }
  \BlankLine	\KwOut{$\texttt{data}$\tcc*{Array of all gathered data, of size $n \cdot p$}
  }
  \BlankLine  $\texttt{data} \leftarrow \texttt{init\_data}$\;
  \For{$i\gets0$ \KwTo $\log_{2}(p)$}{
    $\texttt{size} \leftarrow n \cdot 2^{i}$\;
    $\texttt{dist} \leftarrow 2^{i}$\;
    \uIf {\texttt{id} - \texttt{dist} >= 0}
    {
    send $\texttt{data}[0:\texttt{size}]$ to $\texttt{id} - \texttt{dist}$\;
    }
    \uElse
    {
        send $\texttt{data}[0:\texttt{size}]$ to $\texttt{id} - \texttt{dist} + p$\;
    }
    
    \uIf {\texttt{id} + \texttt{dist} < p}
    {
        receive $\texttt{data}[\texttt{size}:2 \cdot \texttt{size}]$ from $\texttt{id} + \texttt{dist}$\;
    }
    \uElse
    {
        receive $\texttt{data}[\texttt{size}:2 \cdot \texttt{size}]$ from $\texttt{id} + \texttt{dist} - p$\;
    }
  }

  rotate $\texttt{data}$ down by $\texttt{id}$ positions\tcc*{e.g. $\texttt{data}[\texttt{id}] \leftarrow \texttt{data}[0]$}\;
  
    \caption{Bruck Allgather: \texttt{bruck}}\label{alg:bruck}
\end{algorithm2e*}

The Bruck all-gather, as described in Algorithm~\ref{alg:bruck}, minimizes the cost of a final reorder of values, so this algorithm is often used over recursive-doubling or dissemination.  During any given step $i$, process $\texttt{id}$ sends all $\frac{m \cdot 2^{i}}{p}$ values to process $\texttt{id} - 2^{i}$ and receives an additional $\frac{m \cdot 2^{i}}{p}$ values from process $\texttt{id} + 2^{i}$.  This algorithm minimizes the message count so that $\log_{2}(p)$ messages are communicated from each process, while requiring each process to communicate a total of $\frac{m \cdot (p-1)}{p}$ values.  The communication pattern of the Bruck algorithm for Example~\ref{exmp:allgather} is visualized in Figures~\ref{fig:bruck} and~\ref{fig:bruck_data}.

The ring algorithm~\cite{ChanTheory} requires $p-1$ steps during which each process communicates only with neighbors.  At step $i$, each process $\texttt{id}$ sends $\frac{m}{p}$ most recently received values to process $\texttt{id} - 1$ and receives $\frac{m}{p}$ new values from process $\texttt{id} + 1$.  This algorithm requires $p-1$ messages to be communicated per-process while the number of values sent remains $\frac{m \cdot (p-1)}{p}$.

While the Bruck algorithm minimizes the cost of Equation~\ref{eqn:postal}, the ring algorithm optimizes performance in practice for large data sizes $\frac{m}{p}$, likely due to the locality of communication as each step requires communication only among neighbors~\cite{OptimizationMPICH}.  Therefore, the Bruck algorithm is the standard implementation of the MPI\_Allgather for small message sizes, while the ring algorithm is typically used when $\frac{m}{p}$ is large.  The remainder of this paper focuses on further optimizing the Bruck all-gather for small message sizes.

\subsection{Locality-Awareness}
The cost of communication is dependent not only on the message count and number of bytes communicated, but also the relative locations of sending and receiving processes~\cite{BienzEuroMPI}, as noted in the explanation of the ring algorithm.  The cost of communication is also dependent on the number of active processes per node.  When large amounts of data are communicated by many processes at one time, injection bandwidth limits are reached, limiting the speed at which data is transferred~\cite{MaxRate}.  
Modern parallel architectures contain a large number of processes per node.  Often, the on-node cores are further split into multiple CPUs.  Typically, small messages between cores that lie on the same socket, or CPU, are transferred through cache at a much faster rate than larger or inter-socket messages, which are sent through main memory.  However, on-node messages that cross NUMA regions are typically less expensive than those that are injected through the network, with the exception of Spectrum MPI on Power9 machines, where inter-node transfers are significantly less costly than intra-node~\cite{BienzHPEC}.  
\begin{figure*}[ht!]
    \centering
    \includegraphics[width=0.9\textwidth]{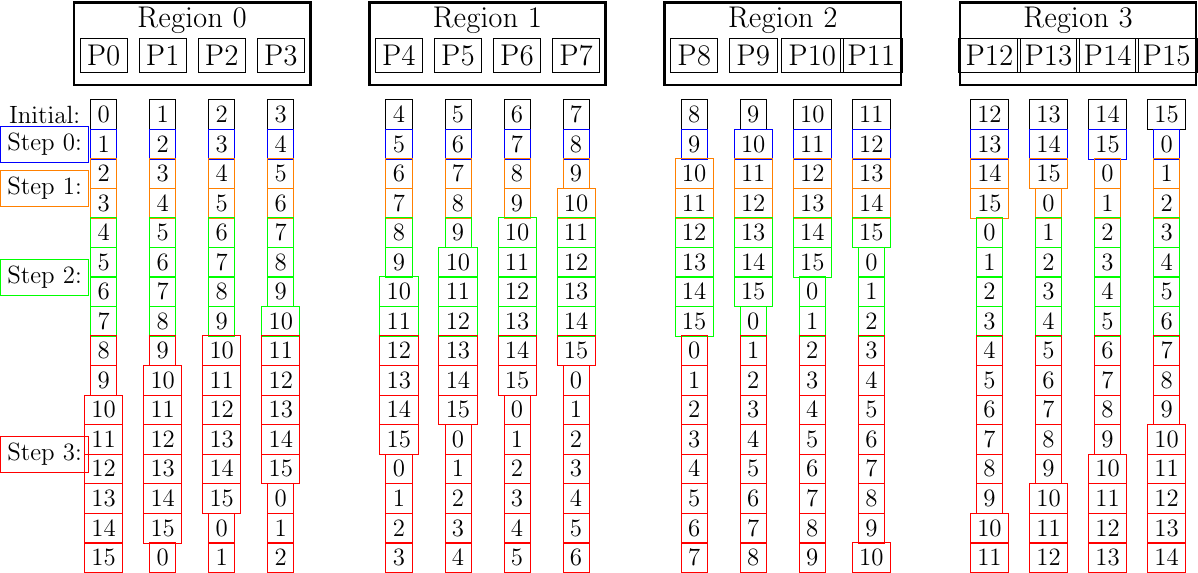}
    \caption{The values from Example~\ref{exmp:allgather} gathered on each process.  The color outlining each value represents the step of the Bruck algorithm during which it was received by the given process.}
    \label{fig:bruck_data}
\end{figure*}

For the remainder of the paper, a \emph{region} of processes describes a group of cores within which communication is inexpensive.  \emph{Non-local} communication is defined as communication between regions, while \emph{local} communication is that within a region.  As an example, a node could be considered a region, with intra-node communication described as local, and inter-node as non-local.

The standard Bruck algorithm is optimized based on the assumption that all messages are equivalent.  As a result, unnecessary non-local communication occurs.  For example, Figure~\ref{fig:bruck} highlights that multiple messages are communicated non-locally between regions.  Step $3$ in the figure requires all processes in region $0$ to send data to a process in region $3$, resulting in multiple non-local messages between the pair of regions.  Furthermore, processes in region $0$ previously sent data to region $3$ in steps $1$ and $2$.  Step $4$ also requires duplicate messages, with each process in region $0$ sending data to a process in region $2$.

The Bruck algorithm also requires single data values to be sent between sets of non-local processes multiple times.  
Figure~\ref{fig:bruck_data} displays the data held by each process after each step of the Bruck all-gather for Example~\ref{exmp:allgather}.  At step $i$, each process sends all data above the data labeled step $i$.  For instance, during step $3$, each process sends the black, blue, and orange data to the corresponding process.  During this step, each process in region $0$ sends the value $3$.  Furthermore, the values $1$, $2$, $4$, and $5$ are all sent in multiple messages originating in this region.  As the processes in region $0$ send to corresponding processes in region $3$ during this step, as shown in Figure~\ref{fig:bruck}, the Bruck algorithm results in not only multiple non-local messages between pairs of regions, but also requires duplicate communication of individual values.  As a result, this algorithm fails to minimize both the number and size of non-local messages. 

\subsection{Related Work}
Hierarchical algorithms reduce injection bandwidth bottlenecks by utilizing a single master process per node.  For example, hierarchical methods for the MPI\_Allgather perform a local gather to a single master process per node, perform a non-local MPI\_Allgather between all master processes, and finally perform a local broadcast from the master process to all other processes per node.  Hierarchical approaches have been used to optimize collective communication in many contexts.  A small subset of these are described in~\cite{HierAllgather, HierCheetah, HierGropp}.  These hierarchical approaches are able to communicate without injection bandwidth bottlenecks.  However, the majority of processes per node sit idle.  Similarly, multi-leader approaches have been explored, particularly with one master process per socket instead of one per-node~\cite{HierMultiLeader}.  These approaches can also communicate without injection bandwidth bottlenecks while utilizing a larger number of processes.  However, the multiple master processes per node once again communicate duplicate non-local messages.  Finally, multi-lane communication has been explored~\cite{multilane}.  This approach utilizes all processes per node so that each communicates a portion of the data.  All inter-node steps are completed before any intra-node communication, reducing the amount of data to be injected into the network.  Multi-lane collective algorithms obtain reduced bandwidth costs.  However, these methods do not reduce the number of steps, or number of inter-node messages beyond the hierarchical approach.

Topology-aware collective algorithms~\cite{TopoAware0,TopoAware1,TopoAware2,TopoAware3,TopoAware4,TopoAware5} optimize the operations for a specific network topology to minimize the number of links traversed during inter-node communication.  For example, a topology-aware implementation of the ring algorithm may map to a single dimension of a 3D torus so that each node shares a link of the network.  While topology-aware collective algorithms improve the cost over standard implementations, they are dependent on the specific topology of a supercomputer and therefore are difficult to port between parallel architectures.
\begin{algorithm2e*}[ht!]
  \DontPrintSemicolon  \KwIn{$\texttt{Comm}$\tcc*{Main MPI Communicator}
        $\texttt{id}$\tcc*{Process ID in $\texttt{Comm}$}
        $p$\tcc*{Number of Processes in $\texttt{Comm}$}
        $\texttt{Comm}_{\ell}$\tcc*{MPI Communicator for local region}
        $\texttt{id}_{\ell}$\tcc*{Process ID in $\texttt{Comm}_{\ell}$}
        $p_{\ell}$\tcc*{Number of Processes in $\texttt{Comm}_{\ell}$}
        $\texttt{r\_n}$\tcc*{Number of regions}
        $\texttt{init\_data}$\tcc*{Initial data to be gathered}
        $n$\tcc*{Number of values in \texttt{init\_data}}
        }
  \BlankLine	\KwOut{$\texttt{data}$\tcc*{Array of all gathered data, of size $n \cdot p$}
  }
  \BlankLine  $\texttt{data} \leftarrow  \texttt{bruck}(\texttt{Comm}_{\ell}, \texttt{id}_{\ell}, p_{\ell}, \texttt{init\_data})$\tcc*{Local gather of initial values}\;
  
  \For{$i\gets0$ \KwTo $\log_{p_{\ell}}(\texttt{r\_n})$}{
    $\texttt{size} \leftarrow n \cdot p_{\ell}^{i+1}$\;
    $\texttt{dist} \leftarrow \texttt{id}_{\ell} * p_{\ell}^{i+1}$\;
    
    \uIf {\texttt{id} - \texttt{dist} $\geq 0$}
    {
        send $\texttt{data}[0:\texttt{size}]$ to $\texttt{id} - \texttt{dist}$\;
    }
    \uElse
    {
        send $\texttt{data}[0:\texttt{size}]$ to $\texttt{id} - \texttt{dist} + p$\;
    }
    
    \uIf {\texttt{id} + \texttt{dist} $< p$}
    {
        receive $\texttt{data}[\texttt{size}:2 \cdot \texttt{size}]$ from $\texttt{id} + \texttt{dist}$\;
    }
    \uElse 
    {
        receive $\texttt{data}[\texttt{size}:2 \cdot \texttt{size}]$ from $\texttt{id} + \texttt{dist} - p$\;
    }
    
    $\texttt{data}[\log_{p_{\ell}}(i) \cdot p_{\ell}] \leftarrow  \texttt{bruck}(\texttt{Comm}_{\ell}, \texttt{id}_{\ell}, p_{\ell}, \texttt{data}[\log_{p_{\ell}}(i) \cdot p_{\ell}])$\tcc*{Local gather of values received in step $i$}\;
  }

    \caption{Locality-Aware Bruck Allgather: \texttt{loc\_bruck}}\label{alg:nap_bruck}
\end{algorithm2e*}

Node-aware optimizations have been introduced for collective algorithms, such as the MPI\_Allreduce operation~\cite{BienzExaMPI}.  While associated performance models show speedup over existing approaches for small all-reduces, overhead of implementing on top of MPI eliminated performance improvements over implementations currently in MPICH.  Locality-aware optimizations also exist for sparse collectives, or irregular communication, such as that required throughout sparse matrix-vector~\cite{BienzJPDC} and sparse matrix-matrix~\cite{BienzIJHPCA} multiplies.

\section{Locality-Aware Bruck Algorithm}~\label{sec:locality}
\begin{figure}[ht!]
    \centering
    \includegraphics[width=0.49\textwidth]{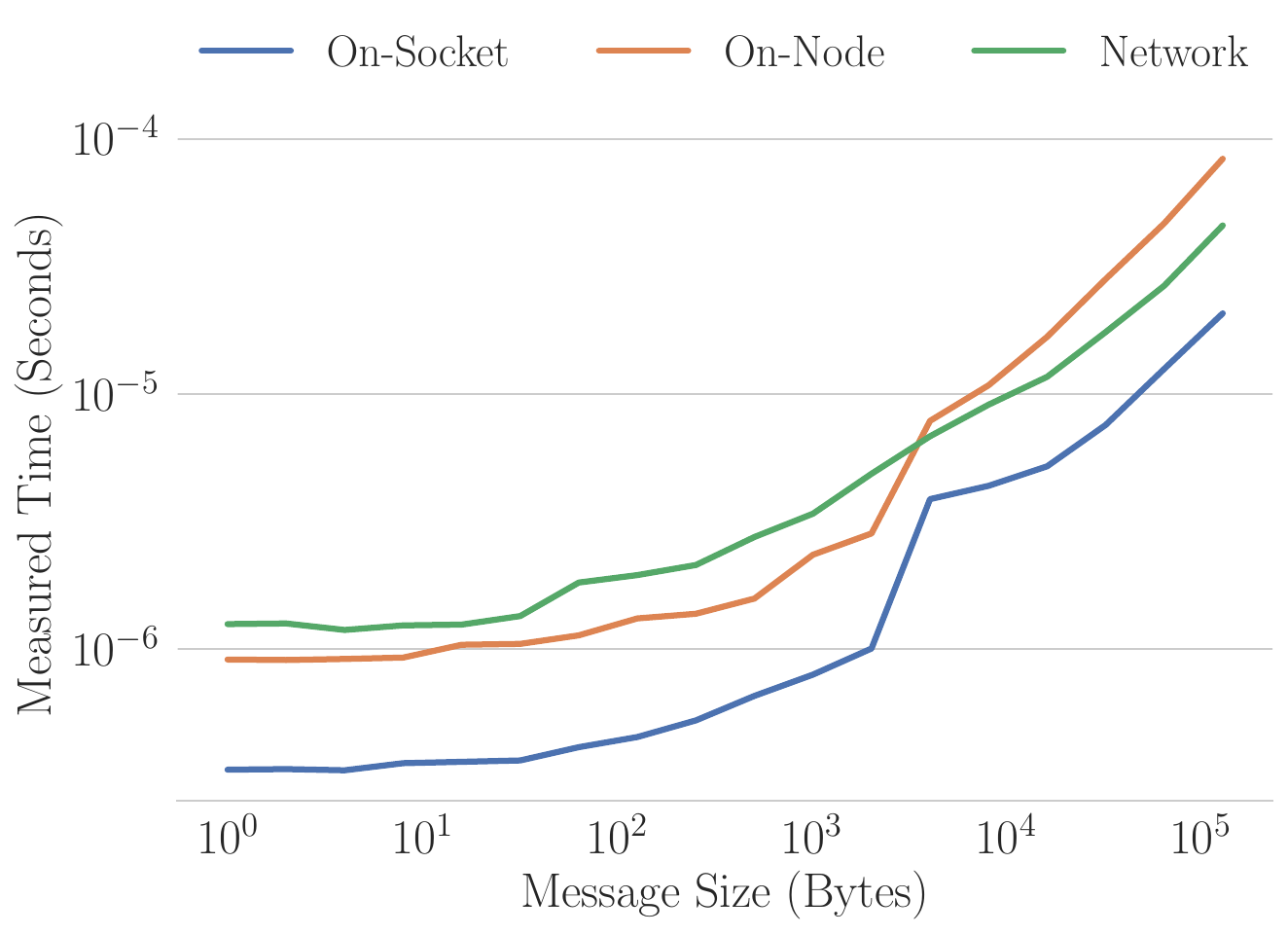}
    \caption{Cost of a single ping-pong of various sizes on Lassen, using Spectrum MPI, split into intra-socket, inter-socket, and inter-node.}
    \label{fig:lassen_cpu_ping_pong}
\end{figure}

State-of-the-art architectures achieve drastic performance differences between intra-socket, inter-socket, and inter-node communication, as shown in Figure~\ref{fig:lassen_cpu_ping_pong}.
The performance model from Equation~\ref{eqn:postal} is improved to account for locality through the following changes
\begin{equation}
    T \leftarrow \alpha_{\ell} \cdot n_{\ell} + \beta_{\ell} \cdot s_{\ell} + \alpha \cdot n + \beta \cdot s 
\end{equation}\label{eqn:locality}
where $\alpha$, $\beta$, $n$, and $s$ are equivalent to the terms in Equation~\ref{eqn:postal}, for non-local communication.  Similarly, $\alpha_{\ell}$, $\beta_{\ell}$, $n_{\ell}$, and $s_{\ell}$ represent the corresponding values for local communication.

On Power9 systems, such as Summit and Lassen, performance is optimized when intra-socket communication is considered local as all other communication is costly.  However, other architectures show notable differences between intra- and inter-node communication, and performance improvements may be shown by treating all intra-node communication as local.

The Bruck algorithm is optimized for locality-awareness as described in Algorithm~\ref{alg:nap_bruck}.  
Similar to existing hierarchical methods, all data is first gathered locally.  However, instead of gathering to a master process, a local all-gather is performed among all processes in each region.  Then, each process within a region sends and receives data with unique regions, before locally gathering all received data.  Note, the first process in each region remains idle during non-local communication to preserve power-of-two exchanges.  During each local all-gather, this process will contribute the original data for simplicity.  Alternatively, an MPI\_Allgatherv operation could be utilized with the first local process contributing no data, or this process could sit idle until all steps of non-local communication have completed.  

\begin{figure*}[ht!]
    \centering
    \includegraphics[width=0.9\textwidth]{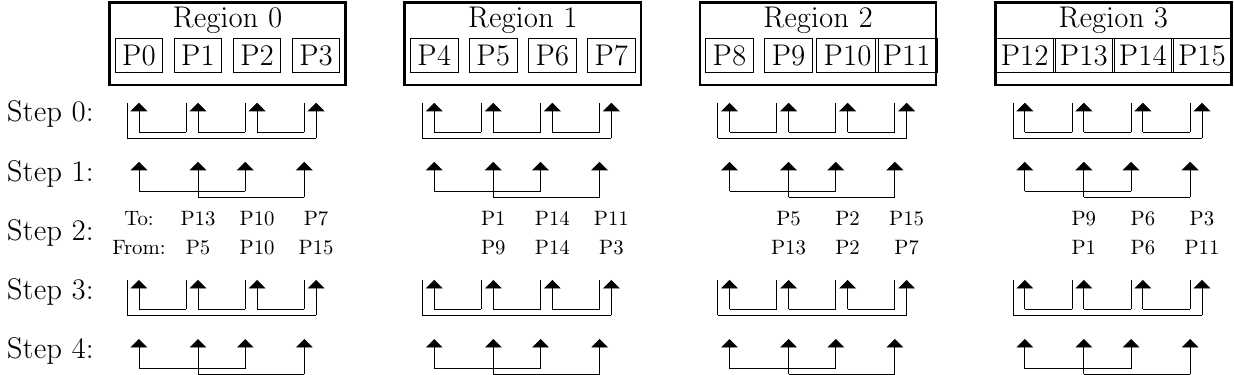}
    \caption{The locality-Aware Bruck algorithm for Example~\ref{exmp:allgather}.  Step $3$ describes all non-local communication, listing the processes to which each sends and from which each receives, for clarity.}
    \label{fig:nap_bruck_pattern}
\end{figure*}

Figure~\ref{fig:nap_bruck_pattern} displays the steps of the locality-aware Bruck algorithm for Example~\ref{exmp:allgather}.
A locality-aware all-gather of Example~\ref{exmp:allgather} requires each process communicate only a single non-local message, compared with the $4$ non-local messages required by the standard Bruck algorithm.

\begin{figure*}[ht!]
    \centering
    \includegraphics[width=0.9\textwidth]{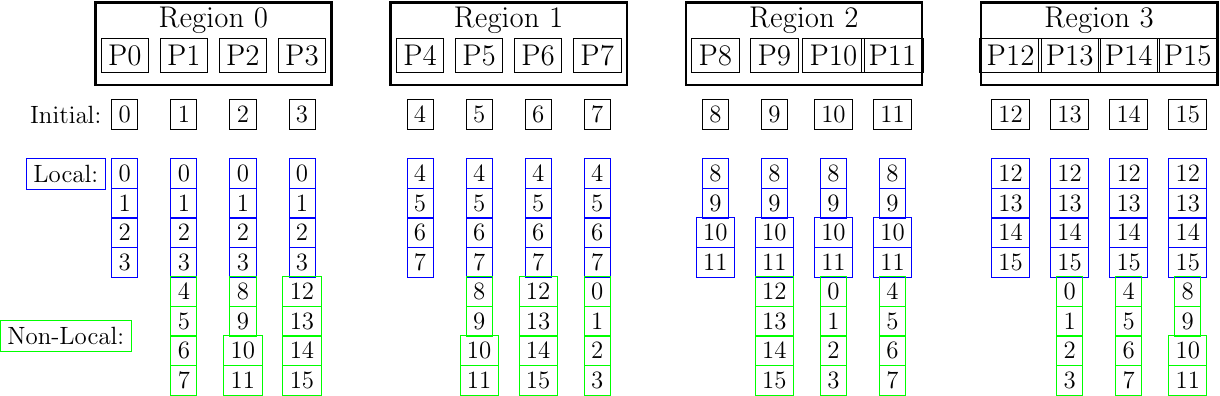}
    \caption{The locality-aware Bruck all-gather for Example~\ref{exmp:allgather}, split into the initial data (black), the data on each process after the local Bruck method (blue), and the new data on each process after the non-local step (green).}
    \label{fig:nap_bruck_data}
\end{figure*}

Figure~\ref{fig:nap_bruck_data} displays data available on each process after the various steps of the locality-aware Bruck algorithm for Example~\ref{exmp:allgather}.  All processes within a region first perform a local all-gather.  Then, each process sends all local data to a unique non-local region, before performing a final all-gather locally.
Each process not only reduces the non-local message count for Example~\ref{exmp:allgather}, but also communicate only $4$ data values non-locally, compared to $15$ non-local values communicated during the standard Bruck algorithm.

\begin{figure*}[ht!]
    \centering
    \includegraphics[width=0.9\textwidth]{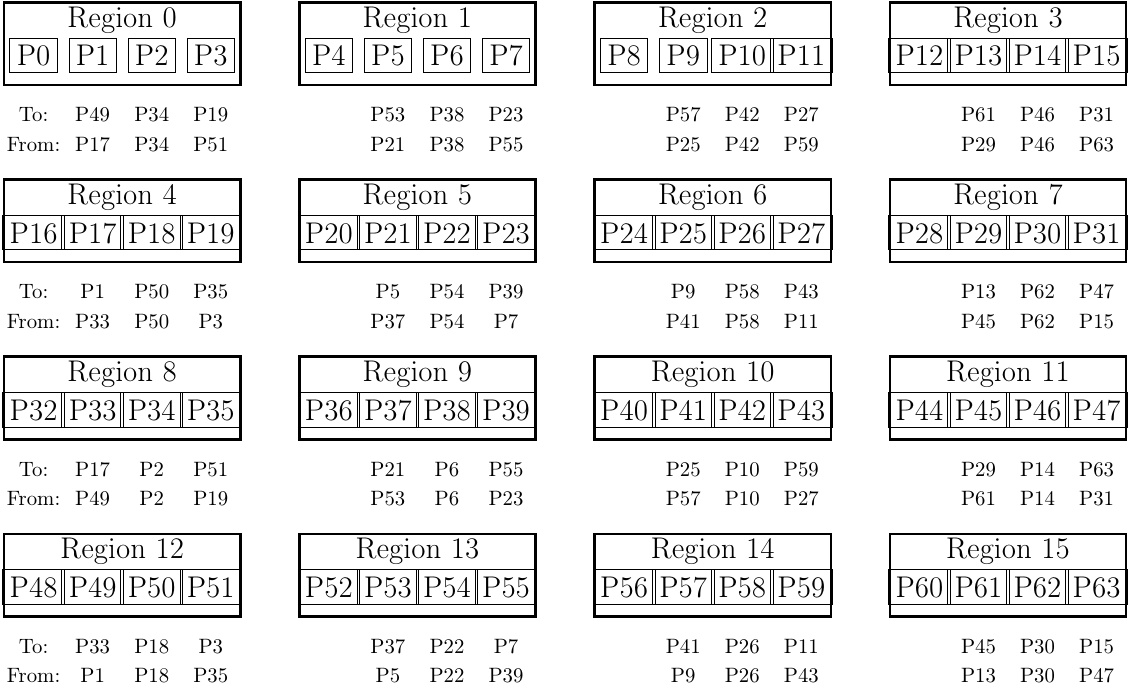}
    \caption{Step 6 of the Locality-Aware Bruck algorithm for $64$ processes distributed across $16$ regions.}
    \label{fig:nap_bruck_pattern_large}
\end{figure*}

The locality-aware Bruck algorithm naturally extends to a larger process count.  Figure~\ref{fig:nap_bruck_pattern_large} shows the additional step of non-local communication required if Example~\ref{exmp:allgather} was extended to $64$ processes with $4$ processes per region.  
Each process holds initial data from a group of four regions, received during the previous steps.  Within the step of non-local communication in Example~\ref{exmp:allgather}, each region now exchanges with the other four groups of regions.  For example, the processes in Region 1 initially hold all data originating within regions $1$ through $4$.  Process $5$ receives data from process $21$, containing all data originating in regions $5$ through $8$.  Process $6$ receives all data originating in regions $9$ through $12$ from process $38$.  Process $7$ receives data from process $55$, which contains all data initially in regions $13$ through $15$ as well as region $0$.

The locality-aware all-gather simply maps to process counts where the number of regions is a power of the number of processes per region.  However, the algorithm naturally extends to any region count.  In the case where the number of regions is not a power of the number of processes per region, a fraction of the processes in each region would sit idle during at least one step of non-local communication.  As a result, an MPI\_Allgatherv would need to be used for the subsequent local step as some processes within the region will hold no new information.

Note, the locality-aware Bruck algorithm naturally extends to additional levels of hierarchy by replacing all calls to \texttt{bruck} in Algorithm~\ref{alg:nap_bruck} with an additional layer of \texttt{loc\_bruck}.  For instance, assume Algorithm~\ref{alg:nap_bruck} performs a node-aware Bruck allgather, with inter-node communication considered non-local.  Instead of calling the standard Bruck allgather in Algorithm~\ref{alg:bruck} on the intra-node communication, Algorithm~\ref{alg:nap_bruck} is used to again to perform a socket-aware allgather on the intra-node communicator.  The standard allgather in Algorithm~\ref{alg:bruck} will then be performed only on intra-socket communicators.

Finally, the locality-aware Bruck algorithm allows for performance reproducibiliy regardless of process placement.  The performance of the standard Bruck algorithm varies with process placement, as the number and size of non-local messages is dependent upon the ordering of the processes.  As locality-aware communication splits the communicators into local and non-local, the ordering of the processes has no impact on non-local communication requirements.

\section{Performance Modeling}\label{sec:perf_models}
The amount of local and non-local communication for both the standard and locality-aware Bruck algorithms can be generalized for any given architecture, process count, and data size.  For a given process count $p$ and data size $\frac{m}{p}$, the standard Bruck algorithm requires $\log_{2}(p)$ non-local messages containing a total of $m-1$ values.  The process with the largest amount of non-local communication requires no local communication.  For example, process $P0$ in Example~\ref{exmp:allgather} sends a total of $15$ values in $4$ non-local messages, while sending no messages locally.  Utilizing the locality-aware performance model from Equation~\ref{eqn:locality}, the modeled cost of the standard Bruck algorithm becomes
\begin{equation}
    T = \log_{2}(p) \cdot \alpha + (b - 1) \beta
\end{equation}\label{eqn:bruck}
where $b$ is the number of bytes in $m$ values.

Assuming $p_{\ell}$ processes per region, there are $r \leftarrow \frac{p}{p_{\ell}}$ regions of processes.  The locality-aware Bruck algorithm requires $\log_{p_{\ell}}(r)$ non-local messages, and $\log_{2}(p_{\ell}) \cdot \left(\log_{p_{\ell}}(r) + 1\right)$ local messages within regions.  During the initial local all-gather, $\frac{p}{m} \cdot (p_{\ell}-1)$ values are communicated locally.  During any given $i^{\texttt{th}}$ step of communication between regions, $\frac{m}{p} \cdot p_{\ell}^{i}$ values are communicated non-locally.  Each following local all-gather requires communication of $\frac{m}{p} \cdot (p_{\ell}^{i+1}-1)$ values within each region.  Therefore, the modeled cost of the locality-aware Bruck algorithm is
\begin{equation}
    T = \log_{p_{\ell}}(r) \cdot \alpha + \frac{b}{p_{\ell}} \beta + \left(\log_{p_{\ell}}(r)+1\right) \cdot \alpha_{\ell} + (b-1) \cdot \beta_{\ell}
\end{equation}\label{eqn:nap_bruck}
where $b$ is the number of bytes in $m$ values.

\begin{figure}[ht!]
    \centering
    \includegraphics[width=0.49\textwidth]{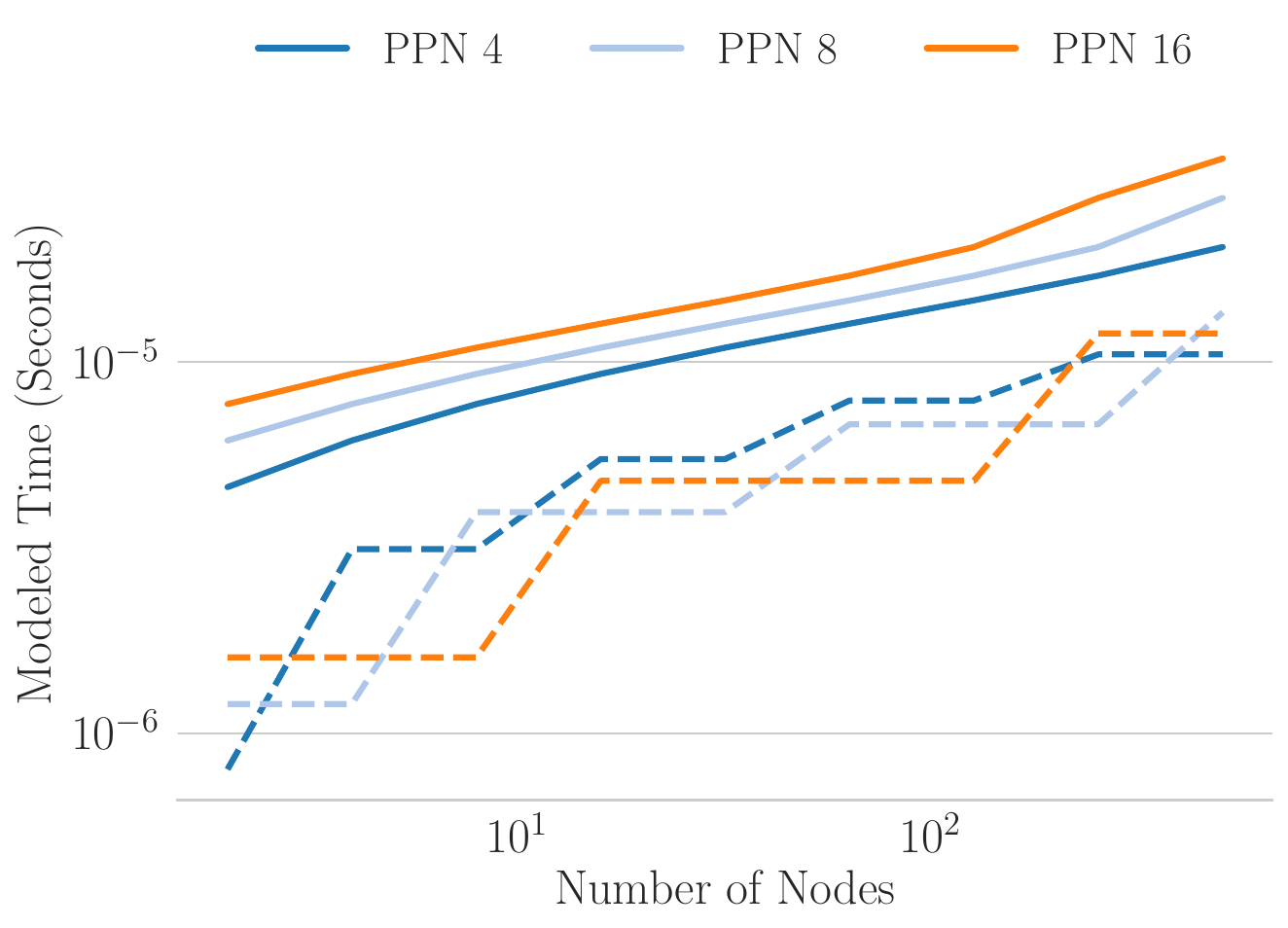}
    \caption{Modeled costs of standard Bruck (solid) vs locality-aware Bruck (dotted) algorithms for various node counts and multiple numbers of processes per node (labeled PPN). The original data size $\frac{m}{p}$ is a single 4-byte integer.}
    \label{fig:model_lassen}
\end{figure}

Figure~\ref{fig:model_lassen} displays the modeled cost of the standard and locality-aware Bruck algorithms on Lassen supercomputer, using the intra-socket and inter-node CPU model parameters from~\cite{BienzHPEC}.  
On Lassen, each socket is considered a separate region due to large costs associated with inter-socket communication.  The model is split into eager and rendezvous protocols, with any message greater than or equal to $8192$ bytes modeled with rendezvous parameters.

The models indicate that the locality-aware allgather outperforms the standard Bruck algorithm for small data sizes.  Furthermore, improvements are amplified with increased numbers of processes per local region.

\begin{figure}[ht!]
    \centering
    \includegraphics[width=0.49\textwidth]{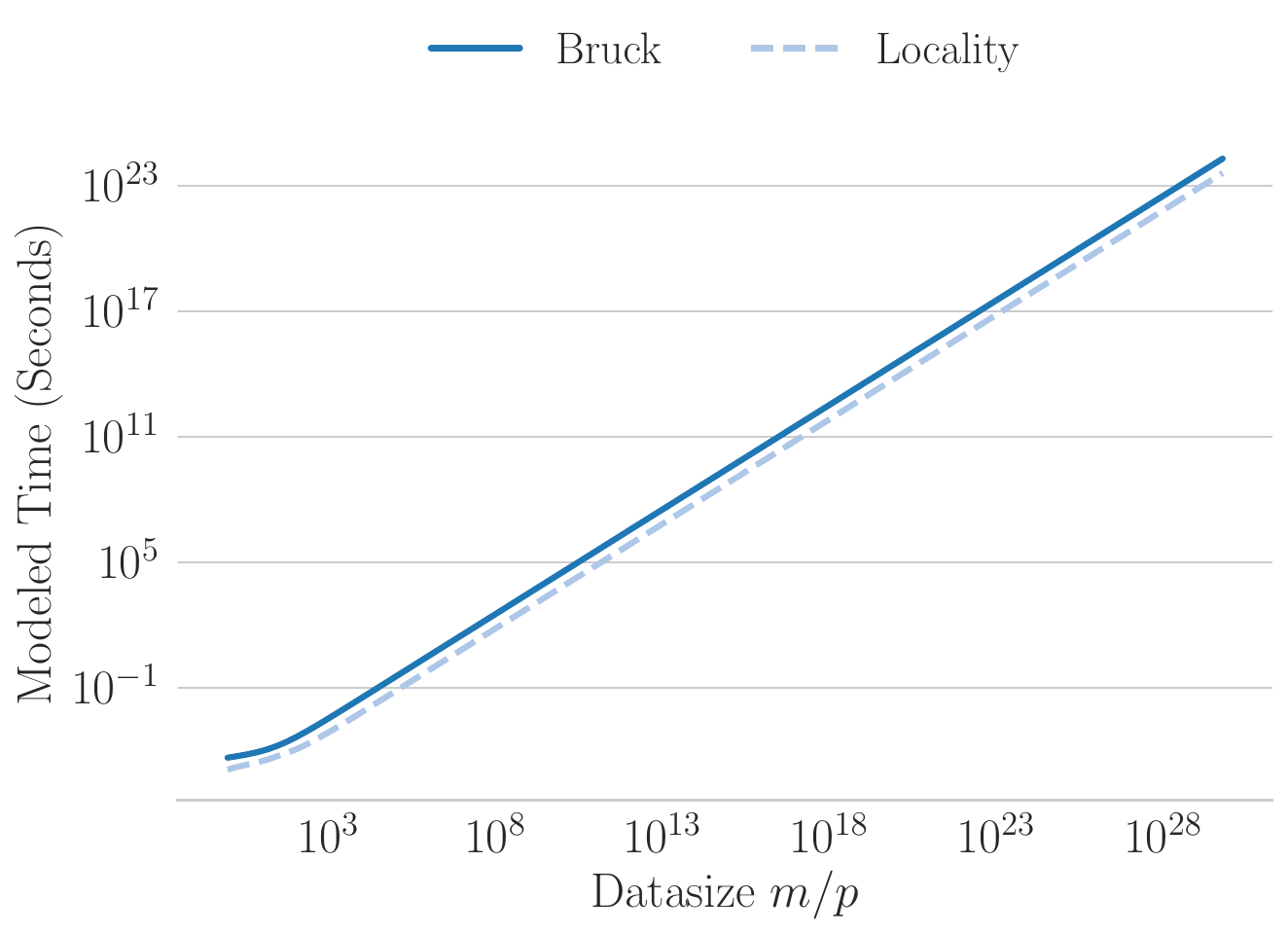}
    \caption{The modeled cost of standard vs locality-aware Bruck algorithms when gathering various data sizes.  The all-gather is performed on $1024$ regions with $16$ processes per region.}
    \label{fig:model_datasize}
\end{figure}

Figure~\ref{fig:model_datasize} displays the modeled costs of standard and locality-aware Bruck all-gathers for a variety of data sizes.  
The size of data has no notable modeled effect on the improvements of the locality-aware Bruck method over the standard algorithm.

\section{Measured Results}\label{sec:results}
The performance of the locality-aware all-gather algorithms was compared against the standard Bruck approach in Algorithm~\ref{alg:bruck}, the hierarchical approach described in~\cite{HierAllgather}, and the multi-lane algorithm from~\cite{multilane} on the following two systems.
\begin{itemize}
    \item \textbf{Quartz:} a system at LLNL with Intel Xeon E5 cores.  The authors consider a node to be a local region on this machine.  Therefore all intra-node communication is considered local, while inter-node communication is non-local. 
    \item \textbf{Lassen:} a Power9 system at LLNL. Only the CPU cores are utilized for performance measurements on this system.  The authors consider a socket, or CPU, to be considered a local region on Lassen.  Therefore, all intra-socket communication is considered local, and all other communication is non-local.  For simplicity, measurements only utilized cores within a single socket per node, so inter-socket but intra-node communication was not included in these measurements.  
\end{itemize}
The data size for each measured all-gather is two $4$-byte integers per process.  Furthermore, all tested local regions were of power-of-2 process counts for simplicity.  Finally, the numbers of regions in each test are a power of the region size.  However, as previously noted, the algorithm would work for other node counts, with fewer processes per region active in at least one step of the algorithm.  All algorithms were implemented by hand, utilizing the \texttt{MPI\_Isend} and \texttt{MPI\_Irecv} operations.  The algorithms are also compared against the implementation within the system install of MPI.

\begin{figure}[ht!]
    \centering
    \includegraphics[width=0.49\textwidth]{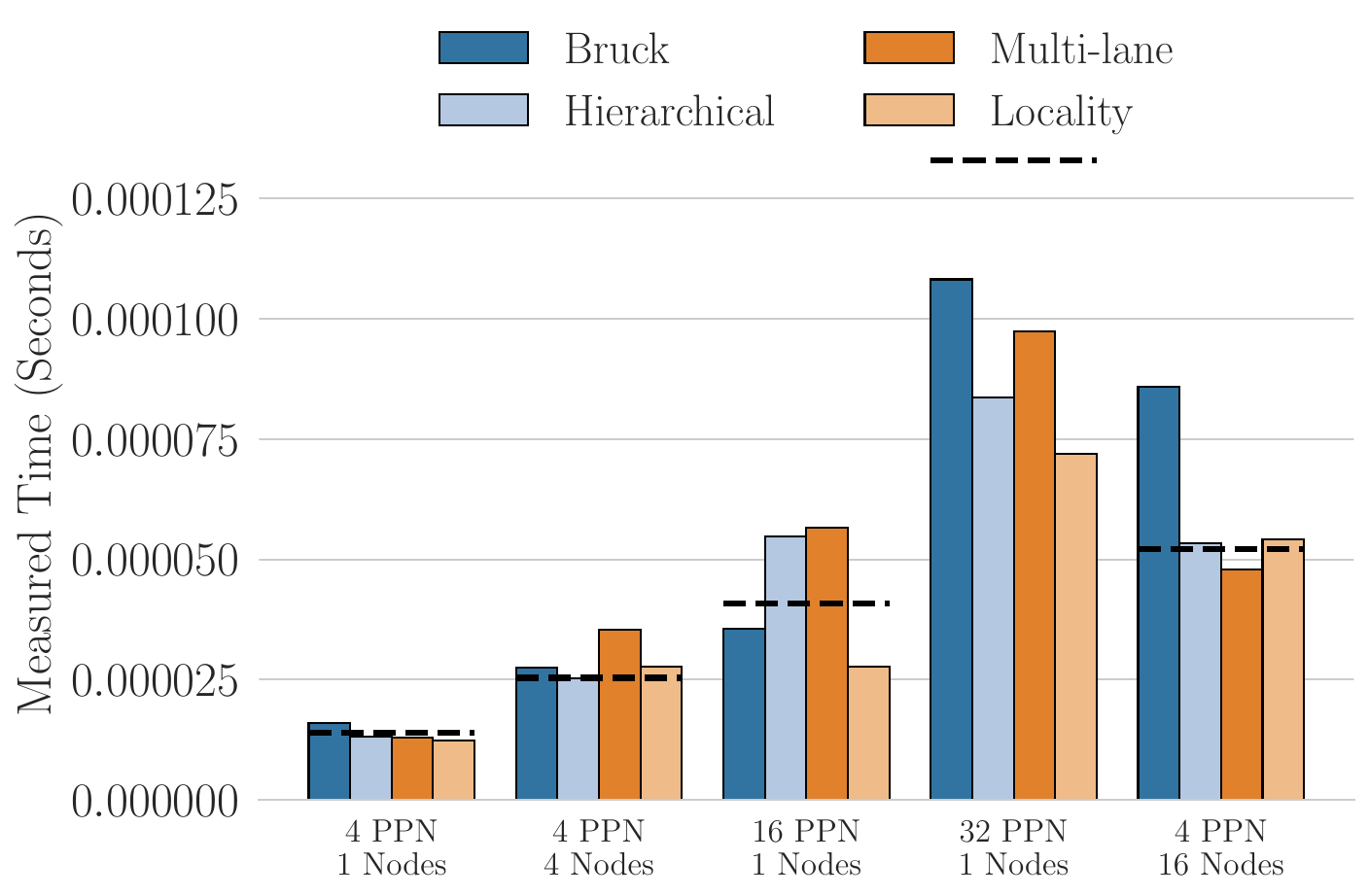}
    \caption{The cost of the standard all-gather method within MVAPICH2 on Quartz compared to the standard Bruck and locality-aware algorithms implemented on top of MPI.  Various numbers of processes within a region, or node (PPN), are tested.}
    \label{fig:quartz}
\end{figure}
The measured costs for the various Bruck algorithms on Quartz are displayed in Figure~\ref{fig:quartz}.  The black dotted line represents the cost of the existing all-gather within MVAPICH2.  Similar to the modeled results, the locality-aware all-gather algorithm improves over the existing Bruck algorithm as well as hierarchical and multi-lane optimizations for many process counts, particularly as the number of processes per local region, or node, increases.  Furthermore, the locality-aware algorithm often improves over the existing implementation within MPI, even though the locality-aware approach incurs overhead from being written on top of MPI.  

\begin{figure}[ht!]
    \centering
    \includegraphics[width=0.49\textwidth]{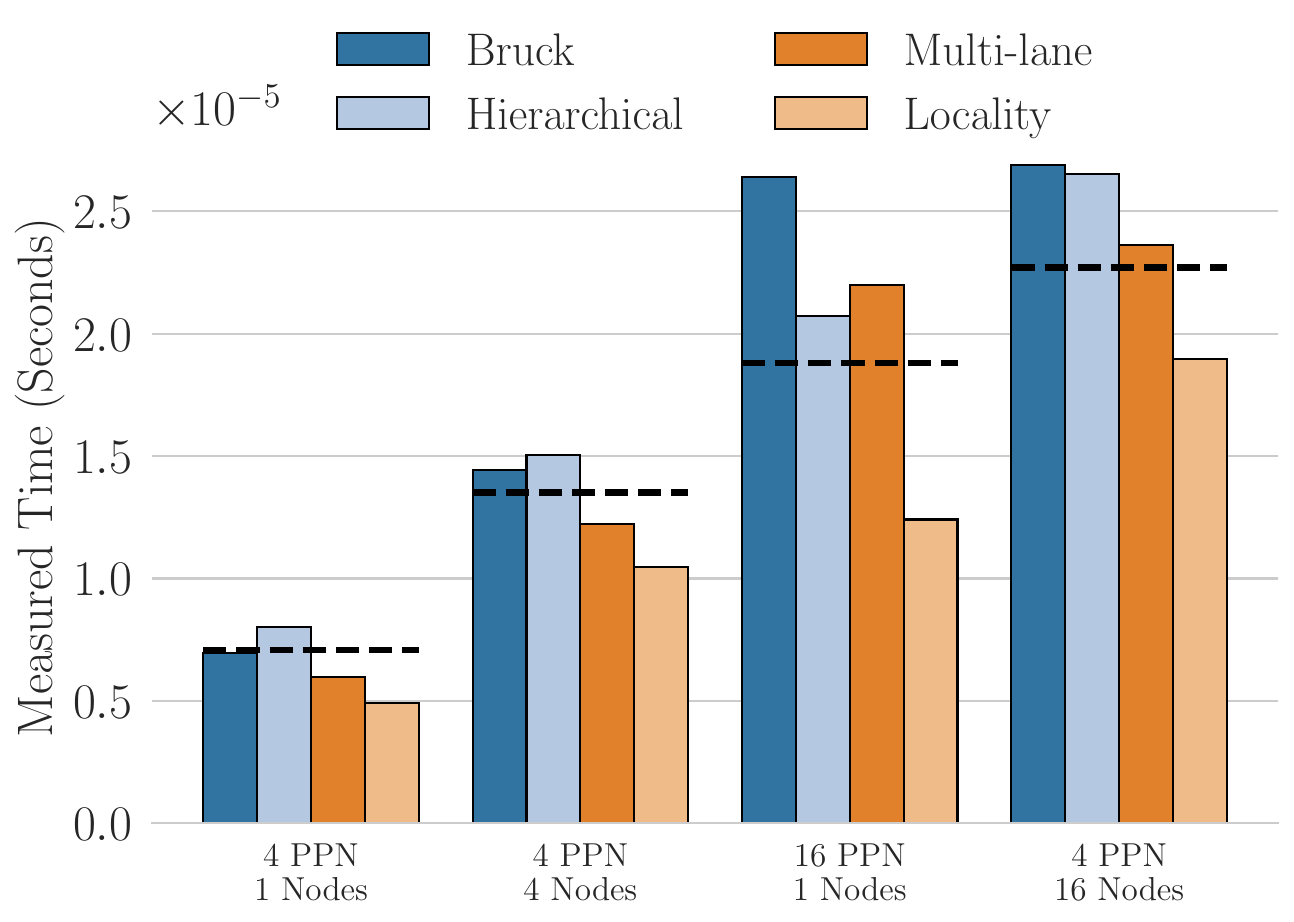}
    \caption{The cost of the various all-gather algorithms implemented on top of MPI, compared to the standard implementation within MPI (black dotted line).  Only a single socket is used per node, so the PPN counts display the number of processes per local region, or socket.}
    \label{fig:lassen}
\end{figure}

The measured costs for the various all-gather algorithms on the CPU cores of Lassen are displayed in Figure~\ref{fig:lassen}.  All algorithms are implemented on top of MPI, and compared to the existing all-gather within Spectrum MPI.
The measurements on Lassen correspond to both the modeled costs and Quartz results.  Locality-aware all-gathers improve over existing methods, and performance improvements are increased with the number of processes per region.  Furthermore, locality-aware optimizations greatly improve over the standard implementation within MPI, even though overhead is incurred as the method is implemented on top of MPI.

\section{Conclusions and Future Work}\label{sec:conclusions}
Standard collective algorithms, such as the all-gather for small data sizes, are optimized to minimize message count and data size.  However, message cost varies with the relative locations of the source and destination processes.  Locality-awareness allows existing algorithms, such as the Bruck all-gather for small data sizes, to be optimized such that non-local, or expensive, communication is minimized.  The locality-aware optimizations for the Bruck algorithm, implemented on top of MPI, improves the performance over existing implementations within MVAPICH2 and Spectrum MPI.  Improvements are amplified as the number of processes per local region increases.

Locality-awareness can be extended to other collectives, removing duplicate non-local messages for small data sizes and reducing the number of non-local bytes to be transported in large collectives.  Furthermore, these algorithms can be optimized for heterogeneous architectures, such as Lassen and Summit, where a large number of CPU cores per GPU typically sit idle.  

\begin{acks}
This material is based in part upon work supported by the Department of Energy under Award Number DE-NA0003966.
\end{acks}

\bibliographystyle{ACM-Reference-Format}
\bibliography{refs}

\end{document}